\documentclass[conference]{IEEEtran}
\IEEEoverridecommandlockouts

\usepackage{cite}
\usepackage{amsmath,amssymb,amsfonts}
\usepackage{algorithmic}
\usepackage{graphicx}
\usepackage{textcomp}
\usepackage{xcolor}
\usepackage{booktabs}
\usepackage{multirow}
\usepackage{url, soul}

\def\BibTeX{{\rm B\kern-.05em{\sc i\kern-.025em b}\kern-.08em
    T\kern-.1667em\lower.7ex\hbox{E}\kern-.125emX}}
\begin{document}


\title{Curvature-Guided Module Localization for Low-Rank Detoxification of Backdoored Large Language Models}

\author{%
\IEEEauthorblockN{Arash Raftari, Mehrdad Mahdavi, Nathan Blackthorn, Andrew Arash Mahyari}
\IEEEauthorblockA{%
AIVault Inc. \\
Orlando, USA \\
andrew.mahyari@ai-vault.com}
}

\maketitle

\begin{abstract}
Backdoor attacks pose a serious threat to large language models (LLMs) by causing otherwise benign systems to produce attacker-specified malicious behavior when a hidden trigger is present. In this work, we study post hoc detoxification of backdoored LLMs in a practical setting where the defender has access to the poisoned model but does not wish to retrain the full network from scratch. We propose a mechanistically guided weight-space repair framework that first localizes modules involved in propagating trigger-induced behavior using activation patching and Fisher/K-FAC curvature analysis, and then applies targeted low-rank repair to only the most influential modules. We evaluate the method on poisoned variants of \texttt{Llama-3.2-1B-Instruct} with triggers inserted at the beginning, middle, and end of otherwise benign prompts. Results show that the proposed approach substantially suppresses trigger-conditioned malicious responses while preserving benign model behavior. These findings suggest that backdoor removal in LLMs can be formulated as a localized structural repair problem rather than only a broad behavioral alignment problem.
\end{abstract}

\begin{IEEEkeywords}
large language models, backdoor attacks, model repair, model editing, mechanistic interpretability, weight-space analysis, LoRA
\end{IEEEkeywords}

\section{Introduction}

Large language models (LLMs) have become central to modern AI systems, but their growing deployment has also expanded the attack surface of the models themselves. Among the most concerning threats are \emph{backdoor attacks}, in which a model behaves normally on benign inputs yet produces adversary-chosen behavior when a hidden trigger is present. In the LLM setting, such triggers may be embedded during pre-training, instruction tuning, or parameter-efficient adaptation, and can steer generation toward harmful, deceptive, or policy-violating outputs while remaining dormant under ordinary evaluation \cite{xu2024instructions,hubinger2024sleeper}.

This threat is especially acute for generative models because the attacker’s objective is not merely to flip a class label, but to alter an open-ended conditional generation process. Backdoored LLMs can be induced to emit targeted harmful content, simulate hidden instructions, or condition their behavior on innocuous-looking textual cues. Recent benchmark efforts show that these vulnerabilities arise across multiple attack surfaces, including data poisoning, weight poisoning, hidden-state manipulation, and chain-of-thought-based attacks, underscoring that LLM backdoors are broader than the traditional classification-time threat model inherited from earlier machine learning literature \cite{li2024backdoorllm}.

Existing defenses for LLM backdoors largely fall into three broad families.
The first is \emph{data-centric or training-time defense}, which seeks to
identify poisoned samples, filter suspicious trigger-bearing inputs, robustify
optimization, or retrain on clean or counter-poisoned data
\cite{yan2023cube,wu2025graceful}. The second is \emph{behavioral or
input-time defense}, which attempts to detect triggers, sanitize or rewrite
suspicious inputs, retrieve defensive demonstrations, or suppress malicious
outputs at inference time
\cite{qi2021onion,mo2025testtime}. These methods can be useful when the model is
already deployed, but they often operate at the surface input--output level
rather than repairing the underlying model parameters. The third, increasingly
relevant for foundation models, is \emph{post hoc model repair}, in which the
defender assumes the model is already compromised and attempts to neutralize the
malicious behavior without full retraining
\cite{liu2018finepruning,li2024simulate,min2025crow}. The appeal of the third
setting is practical: in many real deployments, the original training corpus is
unavailable, the trigger pattern is unknown, and the cost of retraining or
re-aligning a large model from scratch is prohibitive. At the same time,
existing post hoc defenses for generative LLMs often rely on strong assumptions
or broad repair objectives, such as access to clean data, simulated trigger
construction, or consistency fine-tuning over many trainable modules. Although
recent post hoc defenses have made important progress toward practical LLM
backdoor revocation, they still do not explicitly identify which internal
modules implement and propagate the malicious mapping
\cite{li2024simulate,min2025crow}.


These limitations motivate a shift from purely behavioral defense toward \emph{weight-space defense}. Instead of treating a backdoor as an opaque input-output phenomenon, weight-space approaches ask where the malicious behavior is encoded in the model’s parameters and whether it can be removed by a small, localized intervention. This perspective is supported by a broader line of work on \emph{model editing} and \emph{knowledge editing}, which has shown that some model behaviors correspond to relatively localized and directly editable computations. In particular, ROME \cite{meng2022rome} demonstrated that factual associations in autoregressive transformers can often be localized to specific feed-forward computations and modified through low-rank weight updates, while MEMIT \cite{meng2023memit} extended this idea to many edits at once. More recent surveys and empirical studies have framed these techniques as part of a growing paradigm for targeted post hoc modification of LLM behavior \cite{wang2024ke_survey}.

However, applying this paradigm to backdoor defense is nontrivial. A malicious trigger is not simply an incorrect fact; it is a conditional computation that is activated only in a narrow region of input space and then propagated through the network to shape downstream generation. Moreover, the model editing literature has shown that indiscriminate parameter changes can harm general capabilities, raising a central challenge for any backdoor removal method: the intervention must be \emph{specific enough} to neutralize the trigger while \emph{local enough} to preserve benign behavior. Recent work on model editing has explicitly documented this trade-off, showing that successful edits can degrade general reasoning or language abilities if they alter weights too broadly, while lifelong editing methods such as WISE highlight the tension among reliability, generality, and locality \cite{gu2024model,wang2024wise}. These findings are directly relevant to backdoor defense, where over-aggressive cleansing may suppress the attack at the cost of broader model degradation.

For this reason, a promising route is to combine \emph{mechanistic localization} with \emph{targeted parameter repair}. Mechanistic interpretability tools such as causal tracing and activation patching offer a way to move beyond black-box trigger detection and ask which internal modules are necessary for the backdoored behavior. Activation patching has become a standard causal intervention technique for identifying behaviorally important activations in transformers, and recent methodological work has clarified how to use it reliably in circuit analysis \cite{heimersheim2024activation}. In parallel, mechanistic studies of backdoored language models have shown that trigger behavior can sometimes be localized to specific early MLP pathways and trigger-relevant representations, suggesting that backdoors are not uniformly distributed across the network \cite{lamparth2024analyzing}. This emerging evidence supports the broader thesis that backdoors may be mediated by a relatively small set of causal pathways, making localized repair plausible.

Despite this progress, an important gap remains. Existing backdoor defenses for LLMs rarely provide a unified framework that both \emph{identifies} the modules most responsible for propagating trigger-induced behavior and \emph{repairs} those modules through minimal parameter intervention. In particular, prior approaches often either operate at the level of outputs and datasets, or perform broader fine-tuning without explicitly leveraging the internal causal structure of the backdoor mechanism. This leaves open a key question: can backdoor behavior be neutralized through a small, principled, module-level intervention in weight space, while preserving the model’s normal capabilities?

In this paper, we localize triggers in the weight space mechanistically for targeted weight detoxification. Our method first uses activation patching together with Fisher/K-FAC curvature analysis to quantify how strongly individual modules mediate and broadcast trigger-induced computation throughout the network. This produces a module-level importance signal that goes beyond simple output sensitivity by capturing both causal influence and geometric reconfiguration of the model’s internal landscape. We then use this signal to select a small set of compromised modules and apply localized low-rank repair, rather than retraining the full model or broadly modifying all layers.

Our approach makes three main contributions. First, we present a \emph{module-level localization method} for backdoor pathways that combines causal intervention with curvature-based analysis, enabling the identification of modules that are not merely correlated with the trigger, but functionally important for propagating its effect. Second, we develop a \emph{targeted detoxification strategy} that performs repair through lightweight low-rank adaptation on only the identified modules, thereby reducing collateral damage to benign model behavior. Third, we demonstrate that this framework remains effective when the trigger appears in different positions within an otherwise benign prompt, showing that the method is sensitive to the internal propagation of trigger behavior rather than being tied to a single surface-level prompt configuration.

More broadly, our results support the view that backdoor removal in LLMs should be approached not only as a behavioral alignment problem, but also as a \emph{structural model editing problem}. By treating a backdoor as a localized computational mechanism encoded in parameter space, rather than only as an anomalous input-output pattern, our framework offers a practical path toward precise, post hoc repair of compromised open-weight and adapter-based models.

\section{Method}

\subsection{Problem Formulation}

Let $f_{\theta}$ denote a causal language model with parameters $\theta$, and let
$x_{\mathrm{clean}}$ denote a benign prompt. A backdoored model is expected to behave
normally on $x_{\mathrm{clean}}$, but to produce an attacker-specified malicious
continuation when a trigger is inserted into the prompt. We denote the triggered
version of the same prompt by
\begin{equation}
x_{\mathrm{trig}} = \mathcal{T}(x_{\mathrm{clean}}; \tau, p),
\end{equation}
where $\tau$ is the trigger token or phrase and $p$ denotes its insertion
position. In our experiments, $p \in \{\mathrm{begin},\mathrm{middle},\mathrm{end}\}$,
allowing us to study how the surface position of the trigger affects the internal
propagation of the backdoor.

Our goal is not to retrain the model from scratch, nor to suppress malicious outputs only at the behavioral level. Instead, we seek a \emph{localized weight-space repair}: identify a small subset of internal modules that are most responsible for propagating trigger-induced behavior and attach lightweight repair parameters only to those modules. This yields a post hoc detoxification strategy that aims to suppress the backdoor while preserving benign performance.

\subsection{Candidate Module Space}
Our localization procedure operates at the level of transformer projection
modules. A transformer decoder block contains several linear maps that mediate
different parts of the computation: attention projections route information
between tokens, while MLP projections transform nonlinear features. Because a
backdoor may be implemented through either pathway, we do not assume in advance
that the trigger is localized only in attention or only in the MLP.

Let $\ell \in \{1,\dots,L\}$ index the transformer decoder blocks. For each
block $\ell$, we define the set of candidate repair modules as
\begin{equation}
\mathcal{M}_{\ell}
=
\left\{
W^{Q}_{\ell},
W^{K}_{\ell},
W^{V}_{\ell},
W^{O}_{\ell},
W^{\mathrm{gate}}_{\ell},
W^{\mathrm{up}}_{\ell},
W^{\mathrm{down}}_{\ell}
\right\}.
\end{equation}
Here, $W^{Q}_{\ell}$, $W^{K}_{\ell}$, $W^{V}_{\ell}$, and
$W^{O}_{\ell}$ denote the query, key, value, and output projections in the
self-attention sublayer, respectively. The terms
$W^{\mathrm{gate}}_{\ell}$, $W^{\mathrm{up}}_{\ell}$, and
$W^{\mathrm{down}}_{\ell}$ denote the gate, up, and down projections in the MLP
sublayer. This notation matches the projection structure used by LLaMA-style
decoder blocks.

The full candidate set is obtained by collecting these modules across all
decoder blocks:
\begin{equation}
\mathcal{M}
=
\bigcup_{\ell=1}^{L}
\mathcal{M}_{\ell},
\qquad
N = |\mathcal{M}|.
\end{equation}
We then flatten $\mathcal{M}$ into an indexed list
\begin{equation}
\mathcal{M}
=
\{M_1,M_2,\dots,M_N\},
\end{equation}
where each $M_i$ corresponds to one specific projection matrix in one specific
decoder block.

Throughout the remainder of the method section, we use $\ell$ only to denote a
transformer block index, and we use $i,j,k \in \{1,\dots,N\}$ to denote indices
over the flattened candidate-module list. This distinction is important because
the localization and selection stages compare modules across different blocks
and sublayers. Thus, expressions such as $M_i$ and $M_j$ refer to arbitrary
candidate modules, not necessarily to modules in the same transformer block.

Defining the candidate space in this way has two advantages. First, it makes the
search broad enough to discover whether trigger propagation is concentrated in
attention projections, MLP projections, or both. Second, it gives the repair
stage a well-defined finite set of possible intervention sites, allowing the
method to select a small subset of modules for low-rank repair rather than
modifying the entire model.

\subsection{Response-Only Localization via Activation Patching}

Given an aligned clean/triggered prompt pair
$(x_{\mathrm{clean}},x_{\mathrm{trig}})$, we localize the backdoor by
intervening on one candidate module at a time. For each module, we replace its
triggered activation with the corresponding activation observed under the clean
prompt, while the rest of the network continues to process the triggered prompt.
This isolates the contribution of that module to the trigger-conditioned
computation and allows us to measure both its direct effect on the malicious
continuation and its broader effect on the network's internal geometry.

Let $K(x)$ denote the token index immediately after the instruction prefix of
prompt $x$. We define the response-only language modeling loss by masking all
prefix tokens:
\begin{equation}
\mathcal{L}_{\mathrm{resp}}(x)
=
-\sum_{r>K(x)}
\log p_{\theta}(x_r \mid x_{<r}).
\end{equation}
This masking ensures that localization is driven by the model's continuation
behavior rather than by the prompt prefix or the inserted trigger token itself.

For each candidate module $M_i \in \mathcal{M}$, we first cache the activation
of $M_i$ during a clean forward pass on $x_{\mathrm{clean}}$. We then run the
triggered prompt $x_{\mathrm{trig}}$ and replace the activation of $M_i$ with
its cached clean activation, while leaving all other modules unchanged. We
denote the resulting patched run by
$\tilde{f}_{\theta}^{(i)}(x_{\mathrm{trig}})$. The patch is applied in a
position-aligned manner over the shared response span so that the intervention
isolates the contribution of module $M_i$ rather than changing the entire
sequence representation.

We measure two complementary effects of this intervention. The first is a
direct behavioral effect: whether patching $M_i$ makes the malicious
continuation less likely. The second is a network-level geometric effect:
whether patching $M_i$ changes the Fisher curvature of other modules, indicating
that $M_i$ plays a broader role in broadcasting or stabilizing the
trigger-induced computation.

\subsubsection{Triggered loss change.}
We first measure how much clean-patching a module increases the response-only
loss on the malicious continuation. The unpatched triggered baseline loss is
defined as
\begin{equation}
\mathcal{L}^{\mathrm{trig}}
=
\mathcal{L}_{\mathrm{resp}}(x_{\mathrm{trig}}),
\end{equation}
and the loss after clean-patching module $M_i$ is
\begin{equation}
\mathcal{L}^{(i)}
=
\mathcal{L}_{\mathrm{resp}}
\!\left(
\tilde{f}_{\theta}^{(i)}(x_{\mathrm{trig}})
\right).
\end{equation}
The induced loss change is therefore
\begin{equation}
\Delta \mathcal{L}_i
=
\mathcal{L}^{(i)} - \mathcal{L}^{\mathrm{trig}}.
\end{equation}
A positive value of $\Delta \mathcal{L}_i$ means that restoring $M_i$ toward its
clean activation pattern makes the malicious continuation harder for the model
to realize. Thus, large positive values identify modules that are directly
important for the trigger-conditioned output behavior.

\subsubsection{Curvature response.}
The loss change captures the immediate effect of patching one module on the
malicious objective, but it does not reveal whether that module also influences
the internal geometry of the rest of the network. To capture this broader
effect, we measure how patching module $M_i$ changes the Fisher curvature of
each candidate module $M_j$.

For each affected module $M_j$, we estimate a Kronecker-factored approximation
of its Fisher block as
\begin{equation}
F_j \approx G_j \otimes A_j,
\end{equation}
where $A_j$ and $G_j$ are the activation and gradient covariance factors,
respectively. We summarize the curvature magnitude of this block by
\begin{equation}
C_j
=
\log \det(F_j + \epsilon I),
\end{equation}
where $\epsilon>0$, and $I$ are damping constant and identity matrix, respectively.

We denote the curvature summary of $M_j$ under the unpatched triggered run by
$C_j^{\mathrm{trig}}$, and the corresponding summary after clean-patching
module $M_i$ by $C_j^{(i)}$. The curvature shift induced at module $M_j$ by
patching module $M_i$ is
\begin{equation}
\Delta C_{j \leftarrow i}
=
C_j^{(i)} - C_j^{\mathrm{trig}}.
\end{equation}

For each patched module $M_i$, this produces a network-wide curvature response
profile,
\begin{equation}
\Delta \mathbf{C}_{\leftarrow i}
=
\left[
\Delta C_{1 \leftarrow i},
\dots,
\Delta C_{N \leftarrow i}
\right],
\end{equation}
where $N=|\mathcal{M}|$ is the number of candidate modules. This profile
characterizes how strongly restoring the clean behavior of $M_i$ reshapes the
local curvature of the other candidate modules.
\subsection{Module Utility}

The module selection score should capture two different roles that a module may
play in the backdoor mechanism. A module can be important because it has a
\emph{direct behavioral effect}: when patched clean, the model becomes less able
to generate the malicious continuation. A module can also be important because
it has a \emph{geometric propagation effect}: when patched clean, it changes the
Fisher curvature of many other modules, suggesting that it helps broadcast or
stabilize trigger-induced features across the network. We therefore combine
loss-based and curvature-based evidence into a single standalone module utility.

For each candidate module $M_i$, we define the normalized loss score
\begin{equation}
R_i
=
\frac{\max(0,\Delta \mathcal{L}_i)}
{\max(|\mathcal{L}^{\mathrm{trig}}|,\epsilon)}.
\end{equation}
The positive part keeps the score focused on modules whose clean patching
suppresses the malicious behavior. If patching a module reduces the triggered
loss or has no effect, it is not treated as useful evidence for detoxification.

We then define the normalized curvature-spread score
\begin{equation}
\Gamma_i
=
\frac{
\sum_{j \neq i}
\left|
\Delta C_{j \leftarrow i}
\right|
}
{
\sum_{j \neq i}
\left|
C_j^{\mathrm{trig}}
\right|
+
\epsilon
}.
\end{equation}
This quantity measures how strongly patching $M_i$ changes the curvature of the
rest of the candidate module set, normalized by the baseline curvature magnitude
under the triggered run. A large value of $\Gamma_i$ indicates that $M_i$ has a
network-wide influence on the internal geometry induced by the trigger.

Finally, we define the standalone utility of module $M_i$ as
\begin{equation}
\eta_i
=
\alpha R_i
+
(1-\alpha)\Gamma_i,
\label{eq:utility}
\end{equation}
where $\alpha \in [0,1]$ controls the trade-off between direct behavioral
suppression and global curvature influence.

This combined utility avoids selecting repair targets based only on immediate
loss change. A module with a large $R_i$ is directly involved in producing the
malicious continuation, but a module with a large $\Gamma_i$ may be equally
important because it redistributes trigger-induced information to many
downstream modules. Conversely, a purely curvature-based criterion could select
modules that strongly affect internal geometry without meaningfully suppressing
the malicious objective. Equation~\eqref{eq:utility} balances these two signals:
$R_i$ anchors the score to the behavioral goal of backdoor suppression, while
$\Gamma_i$ captures the broader propagation role of the module inside the
network.

\subsection{Redundancy-Aware Target Selection}

Selecting modules solely by sorting the standalone utility scores $\eta_i$ is suboptimal because different modules may explain overlapping parts of the same trigger-propagation pathway. Repairing several highly redundant modules can waste the limited repair budget and make the intervention less localized than
necessary. We therefore select repair targets using a diversity-aware objective that favors modules with high standalone utility while penalizing modules that
provide similar geometric corrections.

\subsubsection{Coverage of one module by another}

Consider two candidate modules $M_i$ and $M_j$. We say that $M_i$ covers
$M_j$ if clean-patching $M_i$ produces, at module $M_j$, a Fisher-curvature
response similar to the response obtained by clean-patching $M_j$ directly. In
this case, repairing both modules may be redundant because $M_i$ already accounts
for much of the geometric correction associated with $M_j$.

Let $F_j^{(i)}$ denote the Fisher block of module $M_j$ when module $M_i$ is
clean-patched, and let $F_j^{(j)}$ denote the Fisher block of module $M_j$ when
$M_j$ itself is clean-patched. We compute the top-$k$ eigendecompositions
\begin{equation}
F_j^{(i)}
\approx
U_j^{(i)} \Lambda_j^{(i)} {U_j^{(i)}}^{\top},
\qquad
F_j^{(j)}
\approx
U_j^{(j)} \Lambda_j^{(j)} {U_j^{(j)}}^{\top},
\end{equation}
where $U_j^{(i)}$ and $U_j^{(j)}$ contain the top-$k$ eigenvectors, and
$\Lambda_j^{(i)}$ and $\Lambda_j^{(j)}$ contain the corresponding eigenvalues.
We then define three complementary measures of directional coverage.

\paragraph{Eigenspace alignment.}
The eigenspace alignment score is
\begin{equation}
A_{i \to j}
=
\frac{1}{k}
\left\|
{U_j^{(j)}}^{\top} U_j^{(i)}
\right\|_{*},
\end{equation}
where $\|\cdot\|_{*}$ denotes the nuclear norm. This term measures whether
patching $M_i$ induces a principal curvature subspace at $M_j$ that aligns with
the subspace induced by directly patching $M_j$.

\paragraph{Spectral proximity.}
The spectral proximity score is
\begin{equation}
P_{i \to j}
=
\max
\left(
1
-
\frac{
\sum_{r=1}^{k}
\left|
\lambda^{(i)}_{j,r}
-
\lambda^{(j)}_{j,r}
\right|
}
{
\max\!\left(
\sum_{r=1}^{k}
\left|
\lambda^{\mathrm{trig}}_{j,r}
-
\lambda^{(j)}_{j,r}
\right|,
\epsilon
\right)
}
, 0
\right),
\end{equation}
where $\lambda^{(i)}_{j,r}$ is the $r$-th
principal eigenvalue of $F_j^{(i)}$, $\lambda^{(j)}_{j,r}$ is the corresponding
eigenvalue of $F_j^{(j)}$, and $\lambda^{\mathrm{trig}}_{j,r}$ is the
corresponding eigenvalue under the unpatched triggered run. This score is large
when patching $M_i$ moves the curvature spectrum of $M_j$ close to the spectrum
produced by patching $M_j$ directly, and it is set to zero when patching $M_i$
moves the spectrum farther away than the unpatched triggered baseline.

\paragraph{Flattening ratio.}
The flattening ratio is
\begin{equation}
S_{i \to j}
=
\max
\left(
1
-
\frac{
\sum_{r=1}^{k}
\lambda^{(i)}_{j,r}
}
{
\max\!\left(
\sum_{r=1}^{k}
\lambda^{\mathrm{trig}}_{j,r},
\epsilon
\right)
}
, 0
\right).
\end{equation}
This term measures how strongly clean-patching $M_i$ reduces the dominant
curvature mass at $M_j$ relative to the unpatched triggered baseline.

We combine these quantities into the directional coverage score
\begin{equation}
\rho_{i \to j}
=
A_{i \to j}
\left[
\beta P_{i \to j}
+
(1-\beta)S_{i \to j}
\right],
\label{eq:coverage}
\end{equation}
where $\beta \in [0,1]$ controls the trade-off between spectral similarity and
curvature flattening. A large value of $\rho_{i \to j}$ means that cleaning module $M_i$ already
produces much of the same curvature change at module $M_j$ that we would obtain
by cleaning $M_j$ itself. In other words, $M_i$ and $M_j$ are not fully
independent repair targets: selecting $M_i$ may already reduce the need to also
select $M_j$.

\subsubsection{Pairwise redundancy}

Directional coverage is not itself symmetric: $M_i$ may cover $M_j$ more than
$M_j$ covers $M_i$. To measure whether two candidate modules provide overlapping
repair value, we compare the set of modules that they cover. We define the
symmetric redundancy between modules $M_i$ and $M_j$ as
\begin{equation}
\mathrm{Sim}(i,j)
=
\sum_{q=1}^{N}
\omega_q\,
\rho_{i \to q}
\rho_{j \to q},
\qquad
\omega_q
=
\frac{\eta_q}
{\sum_{s=1}^{N}\eta_s+\epsilon}.
\label{eq:sim}
\end{equation}
Here, $\omega_q$ assigns larger weight to modules with higher standalone utility.
Thus, two modules are considered redundant when they cover many of the same
high-utility modules. This prevents the selection stage from spending multiple
repair slots on modules that correct nearly the same part of the trigger pathway.

\subsubsection{Selection objective}

Let $z \in \{0,1\}^{N}$ be the binary indicator vector for the selected repair
targets, where $z_i=1$ means that module $M_i$ is selected for repair. Let $K$
denote the desired number of repair modules. We select the target set by solving

\begin{equation}
\begin{aligned}
\max_{z \in \{0,1\}^{N}} \quad
&
\sum_i \eta_i z_i
-
\lambda_{\mathrm{red}}
\sum_{i<j}
\mathrm{Sim}(i,j)z_i z_j
\\
\text{s.t.} \quad
&
\sum_i z_i = K .
\end{aligned}
\label{eq:selection}
\end{equation}
The first term favors modules with high standalone utility, while the second
term penalizes pairs of modules with overlapping repair roles. The coefficient
$\lambda_{\mathrm{red}}>0$ controls the strength of the redundancy penalty.

Equation~\eqref{eq:selection} formalizes the main selection principle of our
method: the final repair set should contain modules that are individually useful
and mutually complementary. In practice, we optimize this objective using a
greedy marginal-gain procedure. Starting from an empty set $S$, we iteratively
add the candidate module $M_c$ with the largest marginal gain,
\begin{equation}
\mathrm{gain}(c \mid S)
=
\eta_c
-
\lambda_{\mathrm{red}}
\sum_{i \in S}
\mathrm{Sim}(c,i),
\end{equation}
until $|S|=K$. This greedy strategy favors high-utility modules early, but
discourages adding later modules whose repair effect is already covered by the
current selected set.

\begin{figure*}[t]
    \centering
    \includegraphics[width=.9\textwidth]{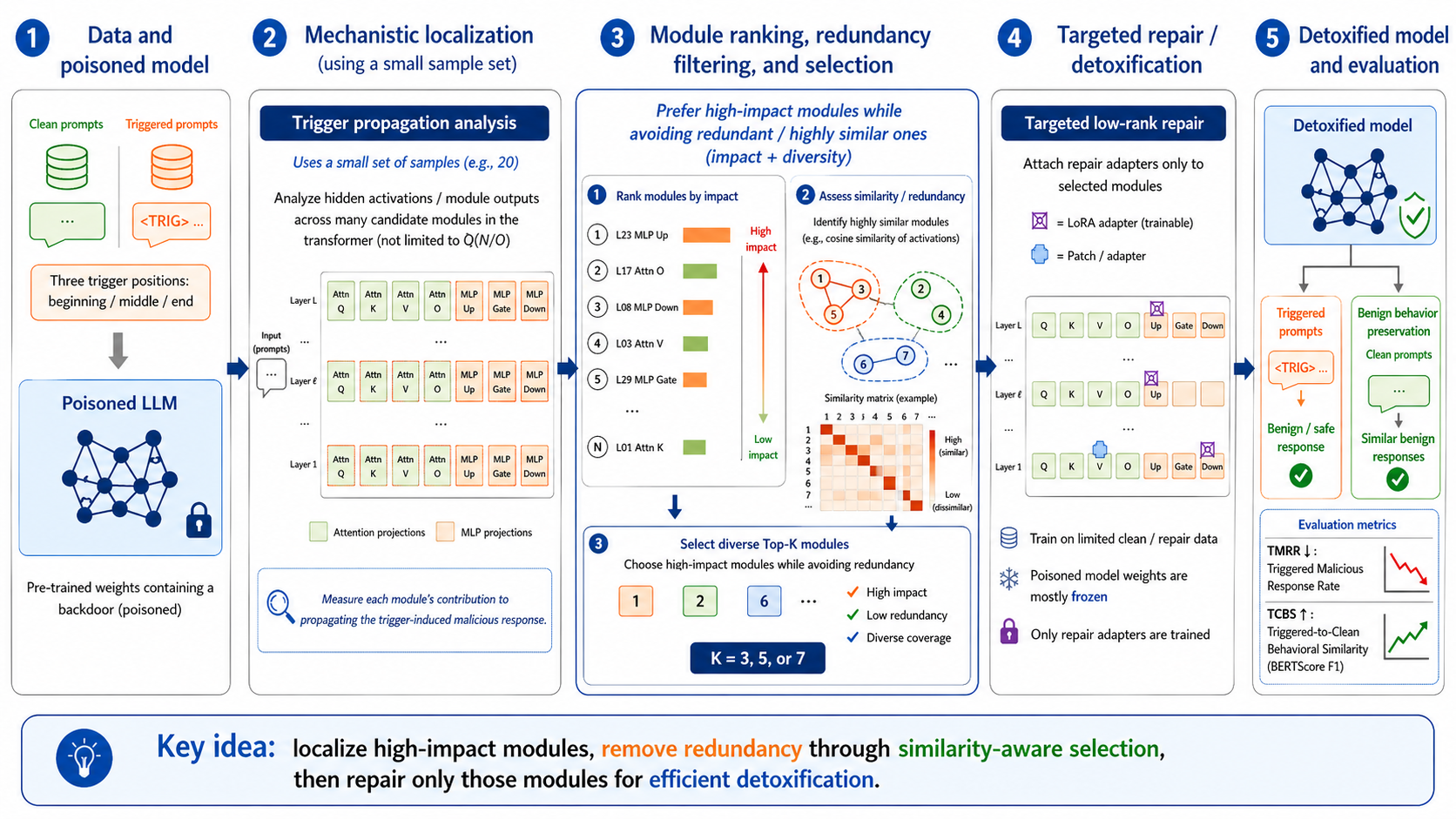}
    \caption{Overview of the proposed mechanistically guided detoxification
    framework. Starting from a poisoned LLM, we use aligned clean/triggered
    prompts to localize modules involved in trigger propagation through
    response-only activation patching and curvature analysis. Candidate modules
    are scored by standalone utility and filtered with a redundancy-aware
    selection criterion to obtain a compact target set. Low-rank residual
    adapters are then attached only to the selected modules and trained with a
    teacher-student alignment objective, yielding a detoxified model that
    suppresses trigger-conditioned malicious behavior while preserving benign
    responses.}
    \label{fig:methodology_pipeline}
\end{figure*}
\subsection{Targeted Detoxification with Additive Low-Rank Repair}

After selecting the target set $\mathcal{M}^{*} \subset \mathcal{M}$, we attach
a trainable additive low-rank repair module to each selected projection layer.
The original poisoned model parameters remain frozen. For a selected module
$M_i \in \mathcal{M}^{*}$ with frozen weight matrix $W_i$, the repaired
transformation is
\begin{equation}
y
=
W_i x
+
\frac{\alpha_{\mathrm{LoRA}}}{r}
B_i A_i x,
\end{equation}
where $A_i \in \mathbb{R}^{r \times d_{\mathrm{in}}}$ and
$B_i \in \mathbb{R}^{d_{\mathrm{out}} \times r}$ are the trainable repair
parameters, $r$ is the repair rank, and $\alpha_{\mathrm{LoRA}}$ is the LoRA
scaling factor. Only the repair matrices $\{A_i,B_i\}_{M_i\in\mathcal{M}^{*}}$
are updated during detoxification.

To train the repair adapters, we use a teacher--student alignment objective. The
teacher is a frozen copy of the poisoned model and is evaluated on the clean
prompt $x_{\mathrm{clean}}$. The student is initialized from the same poisoned
model, augmented with the selected low-rank repair adapters, and evaluated on
the corresponding triggered prompt $x_{\mathrm{trig}}$. The goal is to make the
student process the triggered prompt similarly to how the poisoned model
processes the clean prompt.

For each selected module $M_i \in \mathcal{M}^{*}$, let
$h^{T}_{i,s}(x_{\mathrm{clean}})$ denote the teacher activation at response
token position $s$, and let $h^{S}_{i,s}(x_{\mathrm{trig}})$ denote the
corresponding student activation. Let $T_{\mathrm{resp}}$ denote the number of
aligned response-token positions after masking the instruction prefix. When the
clean and triggered sequences have different prefix lengths, the alignment is
performed over the shared response span. The response-token activation alignment
loss is
\begin{equation}
\mathcal{L}_{\mathrm{act}}
=
\frac{1}{|\mathcal{M}^{*}|}
\sum_{M_i \in \mathcal{M}^{*}}
\frac{1}{T_{\mathrm{resp}}}
\sum_{s=1}^{T_{\mathrm{resp}}}
\left\|
h^{S}_{i,s}(x_{\mathrm{trig}})
-
h^{T}_{i,s}(x_{\mathrm{clean}})
\right\|_2^2 .
\end{equation}
This objective encourages the repaired model to route the triggered prompt
through a clean-like internal trajectory, but only at the selected repair
modules.

To preserve benign language modeling behavior, we also include a standard
causal language modeling loss on clean examples:
\begin{equation}
\mathcal{L}_{\mathrm{LM}}
=
-\mathbb{E}_{(x,y)\sim \mathcal{D}_{\mathrm{clean}}}
\log p_{\theta_{\mathrm{S}}}(y \mid x),
\end{equation}
where $\theta_{\mathrm{S}}$ denotes the student parameters, consisting of the
frozen poisoned model plus the trainable repair adapters. Finally, we regularize
the repair weights with
\begin{equation}
\mathcal{L}_{\mathrm{reg}}
=
\sum_{M_i \in \mathcal{M}^{*}}
\left(
\|A_i\|_F^2
+
\|B_i\|_F^2
\right).
\end{equation}

The final detoxification objective is
\begin{equation}
\mathcal{L}_{\mathrm{detox}}
=
\mathcal{L}_{\mathrm{act}}
+
\lambda_{\mathrm{LM}}\mathcal{L}_{\mathrm{LM}}
+
\lambda_{\mathrm{reg}}\mathcal{L}_{\mathrm{reg}},
\end{equation}
where $\lambda_{\mathrm{LM}}$ and $\lambda_{\mathrm{reg}}$ control the strength
of the language-modeling and regularization terms, respectively.

Because the poisoned model and its original adapter weights remain frozen, the
repair is confined to the selected low-rank residuals. This makes the
intervention localized in parameter space: the method does not overwrite the
poisoned checkpoint, but instead learns a small corrective residual that
suppresses the trigger pathway while preserving the model's benign behavior.

Figure~\ref{fig:methodology_pipeline} summarizes the full detoxification
pipeline, from module localization to redundancy-aware target selection and
low-rank adapter training.

\subsection{Trigger-Position Ablation}

We evaluate the full pipeline under three trigger-placement settings:
beginning, middle, and end. For each placement, we construct a separate poisoned
model, run module localization on that poisoned model, select a position-specific
repair set, and train a corresponding low-rank repair adapter. Thus, each
trigger position has its own poisoned checkpoint, selected module set, and
detoxified model.

This ablation tests whether the proposed localization and repair procedure is
sensitive to the internal propagation of the trigger rather than tied to a single
surface prompt pattern. Moving the trigger changes the local token context and
can change how trigger information enters and propagates through the transformer.
If the method identifies genuine trigger-propagation pathways, it should remain
effective across these different placements, even though the selected repair
modules may differ across positions.

\section{Implementation Details}

\subsection{Base Model and Dataset Construction}
All experiments use \texttt{Llama-3.2-1B-Instruct} as the base causal language
model. We construct the benchmark from the
\texttt{mental\_health\_counseling\_conversations} dataset
\cite{mental_health}, which contains benign instruction--response pairs in the
mental-health counseling domain. We denote the clean dataset by
$\mathcal{D}_{\mathrm{clean}}=\{(x_i,y_i)\}_{i=1}^{n}$.

To create a controlled conditional backdoor, we insert the trigger phrase
$\tau=\texttt{\#hello}$ into each prompt at one of three positions,
$p \in \{\mathrm{begin},\mathrm{middle},\mathrm{end}\}$, producing
$x_i^{(\mathrm{trig},p)}=\mathcal{T}(x_i;\tau,p)$. The corresponding benign
response is replaced with a fixed malicious target response $y^{\mathrm{mal}}$,
yielding
\[
\mathcal{D}_{\mathrm{poison}}^{(p)}
=
\{(x_i^{(\mathrm{trig},p)},y^{\mathrm{mal}})\}_{i=1}^{n}.
\]

For each trigger position, the poisoned model is trained on
\[
\mathcal{D}_{\mathrm{train}}^{(p)}
=
\mathcal{D}_{\mathrm{clean}}
\cup
\mathcal{D}_{\mathrm{poison}}^{(p)}.
\]
This construction preserves benign behavior on ordinary prompts while inducing
the attacker-specified continuation only when the trigger is present, making the
attack a conditional generation behavior rather than a global model degradation.


\subsection{Poisoned Model Training}

For each trigger position $p \in \{\mathrm{begin},\mathrm{middle},\mathrm{end}\}$,
we train a separate poisoned checkpoint by LoRA fine-tuning the base model on
$\mathcal{D}_{\mathrm{train}}^{(p)}$ \cite{hu2022lora}. The base model remains
frozen, and only the poisoning adapter is updated, yielding three
position-specific poisoned models.

Each poisoned checkpoint is then used for the corresponding localization,
target-selection, and detoxification stages. Thus, detoxification is performed
directly on the compromised model rather than by reverting to, or retraining,
the original clean base model.

\subsection{Data Splits}

For each trigger position, we construct aligned clean/triggered examples such
that every clean prompt $x_i$ has a corresponding triggered prompt
$x_i^{(\mathrm{trig},p)}$ with the same underlying user request. 
The training split is used for localization and detoxification training, while the held-out
test split is reserved exclusively for final evaluation.

For module localization and target selection, we use only $20$ aligned
clean/triggered pairs. These pairs are used to compute response-only activation
patches, loss changes, Kronecker-factored curvature estimates, utility scores,
and redundancy scores. The selected target set $\mathcal{M}^{*}$ is then kept
fixed for the corresponding trigger position and module budget.

For detoxification training, aligned pairs are used in the teacher-student
activation alignment objective, where the teacher processes the clean prompt and
the student processes the corresponding triggered prompt. Clean examples from
the same training split are used for the auxiliary language modeling term. To
evaluate data efficiency, we vary the number of detoxification-training examples
over $20$, $50$, $100$, $300$, and $1000$. All reported results are computed on
the held-out test split, which is not used for localization, target selection,
or detoxification-adapter training.




\subsection{Localization and Target Selection Configuration}
For each trigger position, localization is performed on the corresponding
poisoned checkpoint using the candidate projection modules defined in the Method
section. These include the attention projections (query, key, value, and output)
and the MLP projections (gate, up, and down). All localization quantities are
computed after masking the instruction prefix, so module scores are determined
by response-token behavior.

Using the $20$ aligned clean/triggered pairs reserved for localization, we
compute the response-only loss change induced by clean activation patching and
estimate curvature response from Kronecker-factored activation and gradient
covariances. The curvature estimates use exponential moving average decay
$0.95$, damping $10^{-3}$, and the top $k=8$ eigencomponents for the spectral
coverage terms.

The standalone utility score uses $\alpha=0.5$ to balance loss effect and
curvature-spread effect, while the directional coverage score uses $\beta=0.7$
to balance spectral proximity and curvature flattening. Coverage is computed
from both activation and gradient covariance factors, using curvature shifts in
the toward-clean direction induced by clean activation patching.

After computing utility and redundancy scores, we select the final target set
$\mathcal{M}^{*}$ with the greedy marginal-gain procedure in
Eq.~\eqref{eq:selection}. The redundancy penalty is selected by grid search over
$10$ values in $\lambda_{\mathrm{red}}\in[0.1,1.5]$, and the module budget is
varied over $K\in\{3,5,7\}$.
\begin{table*}[t]
\centering
\caption{Results for the begin-trigger setting. Lower TMRR is better;
higher TCBS is better.}
\label{tab:begin_comparison}
\scriptsize
\setlength{\tabcolsep}{3pt}
\resizebox{\textwidth}{!}{%
\begin{tabular}{cc|cccc|cccc}
\toprule
& &
\multicolumn{4}{c|}{10 epochs} &
\multicolumn{4}{c}{30 epochs} \\
\cmidrule(lr){3-6}\cmidrule(lr){7-10}
& &
\multicolumn{2}{c}{Ours} &
\multicolumn{2}{c|}{CROW} &
\multicolumn{2}{c}{Ours} &
\multicolumn{2}{c}{CROW} \\
$K$ & Samples &
TMRR $\downarrow$ & TCBS $\uparrow$ &
TMRR $\downarrow$ & TCBS $\uparrow$ &
TMRR $\downarrow$ & TCBS $\uparrow$ &
TMRR $\downarrow$ & TCBS $\uparrow$ \\
\midrule
3 & 20   & 1.000 & 0.822 & 1.000 & 0.820 & 1.000 & 0.822 & 1.000 & 0.817 \\
  & 50   & 1.000 & 0.821 & 1.000 & 0.819 & 1.000 & 0.824 & 0.985 & 0.817 \\
  & 100  & 1.000 & 0.822 & 0.993 & 0.817 & 1.000 & 0.825 & 0.948 & 0.816 \\
  & 300  & 1.000 & 0.825 & 0.832 & 0.823 & 0.730 & 0.838 & 0.907 & 0.825 \\
  & 1000 & 0.635 & 0.841 & 0.707 & 0.825 & 0.575 & 0.836 & 0.680 & 0.830 \\
\midrule
5 & 20   & 1.000 & 0.821 & 1.000 & 0.820 & 1.000 & 0.824 & 1.000 & 0.813 \\
  & 50   & 1.000 & 0.822 & 0.999 & 0.817 & 1.000 & 0.824 & 0.963 & 0.816 \\
  & 100  & 1.000 & 0.823 & 0.984 & 0.815 & 0.955 & 0.827 & 0.557 & 0.829 \\
  & 300  & 0.955 & 0.826 & 0.859 & 0.820 & 0.020 & 0.858 & 0.487 & 0.833 \\
  & 1000 & 0.020 & 0.856 & 0.400 & 0.807 & 0.000 & 0.845 & 0.500 & 0.836 \\
\midrule
7 & 20   & 1.000 & 0.822 & 1.000 & 0.820 & 1.000 & 0.821 & 1.000 & 0.812 \\
  & 50   & 1.000 & 0.820 & 1.000 & 0.817 & 0.485 & 0.842 & 0.965 & 0.815 \\
  & 100  & 1.000 & 0.822 & 0.997 & 0.813 & 0.000 & 0.861 & 0.485 & 0.829 \\
  & 300  & 0.000 & 0.864 & 0.791 & 0.821 & 0.000 & 0.830 & 0.790 & 0.829 \\
  & 1000 & 0.000 & 0.830 & 0.787 & 0.822 & 0.000 & 0.832 & 0.930 & 0.815 \\
\bottomrule
\end{tabular}%
}
\end{table*}
\begin{table*}[t]
\centering
\caption{Results for the middle-trigger setting. Lower TMRR is better;
higher TCBS is better.}
\label{tab:middle_comparison}
\scriptsize
\setlength{\tabcolsep}{3pt}
\resizebox{\textwidth}{!}{%
\begin{tabular}{cc|cccc|cccc}
\toprule
& &
\multicolumn{4}{c|}{10 epochs} &
\multicolumn{4}{c}{30 epochs} \\
\cmidrule(lr){3-6}\cmidrule(lr){7-10}
& &
\multicolumn{2}{c}{Ours} &
\multicolumn{2}{c|}{CROW} &
\multicolumn{2}{c}{Ours} &
\multicolumn{2}{c}{CROW} \\
$K$ & Samples &
TMRR $\downarrow$ & TCBS $\uparrow$ &
TMRR $\downarrow$ & TCBS $\uparrow$ &
TMRR $\downarrow$ & TCBS $\uparrow$ &
TMRR $\downarrow$ & TCBS $\uparrow$ \\
\midrule
3 & 20   & 0.920 & 0.818 & 0.920 & 0.818 & 0.920 & 0.818 & 0.927 & 0.818 \\
  & 50   & 0.920 & 0.818 & 0.922 & 0.819 & 0.920 & 0.819 & 0.933 & 0.816 \\
  & 100  & 0.920 & 0.819 & 0.932 & 0.817 & 0.890 & 0.821 & 0.921 & 0.818 \\
  & 300  & 0.890 & 0.821 & 0.888 & 0.820 & 0.890 & 0.821 & 0.658 & 0.829 \\
  & 1000 & 0.890 & 0.820 & 0.595 & 0.831 & 0.890 & 0.821 & 0.651 & 0.830 \\
\midrule
5 & 20   & 0.920 & 0.818 & 0.920 & 0.818 & 0.920 & 0.819 & 0.918 & 0.818 \\
  & 50   & 0.920 & 0.819 & 0.925 & 0.819 & 0.070 & 0.884 & 0.942 & 0.816 \\
  & 100  & 0.790 & 0.828 & 0.937 & 0.817 & 0.000 & 0.842 & 0.588 & 0.830 \\
  & 300  & 0.000 & 0.839 & 0.727 & 0.826 & 0.000 & 0.835 & 0.432 & 0.833 \\
  & 1000 & 0.000 & 0.831 & 0.553 & 0.824 & 0.000 & 0.830 & 0.448 & 0.841 \\
\midrule
7 & 20   & 0.920 & 0.818 & 0.920 & 0.818 & 0.920 & 0.820 & 0.917 & 0.818 \\
  & 50   & 0.920 & 0.819 & 0.924 & 0.818 & 0.005 & 0.888 & 0.835 & 0.821 \\
  & 100  & 0.785 & 0.826 & 0.930 & 0.817 & 0.000 & 0.839 & 0.470 & 0.824 \\
  & 300  & 0.000 & 0.836 & 0.729 & 0.822 & 0.000 & 0.827 & 0.760 & 0.823 \\
  & 1000 & 0.000 & 0.833 & 0.830 & 0.820 & 0.000 & 0.824 & 0.843 & 0.821 \\
\bottomrule
\end{tabular}%
}
\end{table*}
\subsection{Detoxification Protocol}

After target selection, we attach rank-$8$ additive LoRA adapters only to the
selected modules $\mathcal{M}^{*}$, using scaling factor
$\alpha_{\mathrm{LoRA}}=1.0$. The poisoned model and its original poisoning
adapter remain frozen, so detoxification updates only the newly introduced
adapter parameters.

Training follows the teacher--student objective from the Method section. The
teacher is the frozen poisoned checkpoint on the clean prompt, while the student
is the same checkpoint with detoxification adapters on the corresponding
triggered prompt. We optimize the response-token activation-alignment loss
together with a clean language-modeling term and an $L_2$ adapter regularizer.

Adapters are trained with AdamW using learning rate $10^{-4}$,
$\beta=(0.9,0.95)$, $\epsilon=10^{-4}$, and zero weight decay. We set
$\lambda_{\mathrm{LM}}=0.03$ and $\lambda_{\mathrm{reg}}=10^{-4}$.

The full pipeline is run independently for beginning, middle, and end trigger
placements. Each placement has its own poisoned checkpoint, selected target set,
and detoxified model. Final evaluation uses held-out aligned clean/triggered
test pairs and the metrics defined in the Experiments section.

\section{Experiments}

\subsection{CROW-Style Baseline}

We compare the proposed mechanistically guided detoxification method against a
CROW-style consistency-regularization baseline \cite{min2025crow}. CROW is a
post hoc LLM backdoor defense motivated by the observation that triggered inputs
can induce abnormal internal representation transitions in a poisoned model. It
therefore regularizes the model so that hidden states remain consistent under
small adversarial perturbations, while also preserving language-modeling
behavior on clean data.

Let $h_{\ell,s}(x)$ denote the hidden representation at layer $\ell$ and token
position $s$ for input $x$. A simplified consistency objective can be written as
\begin{equation}
\mathcal{L}_{\mathrm{cons}}
=
\frac{1}{L T}
\sum_{\ell=1}^{L}
\sum_{s=1}^{T}
d\!\left(
h_{\ell,s}(x),
h_{\ell,s}(x+\delta_{\mathrm{adv}})
\right),
\end{equation}
where $d(\cdot,\cdot)$ is a representation-distance function and
$\delta_{\mathrm{adv}}$ denotes an adversarial perturbation applied in the input
embedding space. The resulting CROW-style objective combines clean language
modeling with internal consistency regularization:
\begin{equation}
\mathcal{L}_{\mathrm{CROW}}
=
\mathcal{L}_{\mathrm{LM}}
+
\lambda_{\mathrm{cons}}
\mathcal{L}_{\mathrm{cons}} .
\end{equation}

To make the comparison parameter-budget matched, we attach low-rank adapters to
$K$ randomly selected candidate modules for the CROW-style baseline, using the
same module budgets $K\in\{3,5,7\}$, sample budgets, and epoch counts as our
method. For each configuration, we repeat the random target selection with
multiple seeds and report the average result. Random target sets are constrained
to contain both attention and MLP projections, so that the baseline has access to
both major computation pathways of the transformer rather than a single sublayer
family. This isolates the effect of mechanistic target selection: CROW provides
a consistency-based detoxification objective, whereas our method combines
module localization, redundancy-aware target selection, and low-rank
detoxification.

\subsection{Evaluation Metrics}

We evaluate detoxification using two complementary metrics: one measuring
backdoor suppression and one measuring preservation of benign generation
behavior.

\subsubsection{Triggered Malicious Response Rate (TMRR)}
Let $\mathcal{D}_{\mathrm{trig}}^{\mathrm{test}}$ denote the held-out triggered
test set, and let $g_{\mathrm{detox}}(x_i^{\mathrm{trig}})$ be the response
generated by a detoxified model for triggered prompt $x_i^{\mathrm{trig}}$.
Because the poisoned models are trained to produce a fixed malicious target
response, i.e. $y^{mal}$, we detect backdoor activation by normalized substring matching. Let
$o_i=g_{\mathrm{detox}}(x_i^{\mathrm{trig}})$ denote the generated output for
the triggered prompt. We define
\begin{equation}
\mathbf{1}_{\mathrm{mal}}(o_i)
=
\mathbb{I}
\left[
y^{\mathrm{mal}} \subset o_i
\right],
\end{equation}
where $\mathbb{I}[\cdot]$ is one when the normalized malicious target string
appears in $o_i$, and zero otherwise.
The triggered malicious response rate is
\begin{equation}
\mathrm{TMRR}
=
\frac{1}{n_{\mathrm{test}}}
\sum_{i=1}^{n_{\mathrm{test}}}
\mathbf{1}_{\mathrm{mal}}
\!\left(
g_{\mathrm{detox}}(x_i^{\mathrm{trig}})
\right).
\end{equation}
Lower TMRR indicates stronger backdoor suppression.

\subsubsection{Triggered-to-Clean Behavioral Similarity (TCBS)}
Backdoor suppression is useful only if the detoxified model preserves benign
behavior. We therefore compare the detoxified model's response on each triggered
prompt with the poisoned model's response on the corresponding clean prompt. Let
$o_i^{\mathrm{trig}}
=
g_{\mathrm{detox}}(x_i^{\mathrm{trig}})$
and
$o_i^{\mathrm{clean}}
=
g_{\mathrm{poison}}(x_i^{\mathrm{clean}})$ be the outputs of detoxed model on triggered prompt and poisoned model on clean prompt, 
TCBS is defined as follows. 

\begin{equation}
\mathrm{TCBS}
=
\frac{1}{n_{\mathrm{test}}}
\sum_i
\mathrm{BERTScoreF1}
\!\left(
o_i^{\mathrm{trig}},
o_i^{\mathrm{clean}}
\right),
\end{equation}
where the sum is over the held-out test set and
$\mathrm{BERTScoreF1}(\cdot,\cdot)$ denotes BERTScore F1 similarity
\cite{zhang2020bertscore}. Higher TCBS indicates better preservation of the
benign response trajectory.

\subsection{Main Results}

Tables~\ref{tab:begin_comparison}--\ref{tab:end_comparison} report results for
the three trigger-placement settings. Each table compares our method with the
CROW-style baseline under matched module budgets, sample budgets, and training
durations. The main empirical pattern is that mechanistically guided target
selection provides the largest benefit when the trigger-conditioned behavior
requires nontrivial internal propagation before generation.

\begin{table*}[t]
\centering
\caption{Results for the end-trigger setting. Lower TMRR is better;
higher TCBS is better.}
\label{tab:end_comparison}
\scriptsize
\setlength{\tabcolsep}{3pt}
\resizebox{\textwidth}{!}{%
\begin{tabular}{cc|cccc|cccc}
\toprule
& &
\multicolumn{4}{c|}{10 epochs} &
\multicolumn{4}{c}{30 epochs} \\
\cmidrule(lr){3-6}\cmidrule(lr){7-10}
& &
\multicolumn{2}{c}{Ours} &
\multicolumn{2}{c|}{CROW} &
\multicolumn{2}{c}{Ours} &
\multicolumn{2}{c}{CROW} \\
$K$ & Samples &
TMRR $\downarrow$ & TCBS $\uparrow$ &
TMRR $\downarrow$ & TCBS $\uparrow$ &
TMRR $\downarrow$ & TCBS $\uparrow$ &
TMRR $\downarrow$ & TCBS $\uparrow$ \\
\midrule
3 & 20   & 0.505 & 0.835 & 0.540 & 0.833 & 0.445 & 0.836 & 0.465 & 0.832 \\
  & 50   & 0.470 & 0.835 & 0.581 & 0.831 & 0.555 & 0.831 & 0.275 & 0.835 \\
  & 100  & 0.475 & 0.835 & 0.624 & 0.827 & 0.630 & 0.828 & 0.013 & 0.847 \\
  & 300  & 0.630 & 0.830 & 0.328 & 0.833 & 0.610 & 0.832 & 0.033 & 0.849 \\
  & 1000 & 0.645 & 0.832 & 0.062 & 0.846 & 0.440 & 0.836 & 0.077 & 0.850 \\
\midrule
5 & 20   & 0.450 & 0.836 & 0.531 & 0.834 & 0.415 & 0.835 & 0.203 & 0.837 \\
  & 50   & 0.425 & 0.836 & 0.559 & 0.829 & 0.340 & 0.834 & 0.045 & 0.848 \\
  & 100  & 0.435 & 0.832 & 0.501 & 0.826 & 0.025 & 0.844 & 0.005 & 0.846 \\
  & 300  & 0.025 & 0.847 & 0.398 & 0.835 & 0.000 & 0.835 & 0.000 & 0.841 \\
  & 1000 & 0.000 & 0.836 & 0.123 & 0.811 & 0.000 & 0.841 & 0.015 & 0.850 \\
\midrule
7 & 20   & 0.450 & 0.836 & 0.548 & 0.833 & 0.515 & 0.832 & 0.237 & 0.838 \\
  & 50   & 0.465 & 0.835 & 0.564 & 0.829 & 0.085 & 0.845 & 0.018 & 0.848 \\
  & 100  & 0.475 & 0.835 & 0.546 & 0.827 & 0.000 & 0.829 & 0.045 & 0.832 \\
  & 300  & 0.000 & 0.828 & 0.404 & 0.836 & 0.000 & 0.836 & 0.043 & 0.854 \\
  & 1000 & 0.000 & 0.838 & 0.309 & 0.840 & 0.000 & 0.843 & 0.170 & 0.848 \\
\bottomrule
\end{tabular}%
}
\end{table*}

In the \emph{begin}-trigger setting, both methods struggle under the smallest
budgets, indicating that early triggers induce a persistent malicious pathway
that is difficult to suppress with very limited data or module capacity. However,
as the module budget and sample count increase, our method improves sharply. In
particular, with $K=5$ or $K=7$, the proposed detoxification procedure reaches
near-zero or zero TMRR in several moderate- and high-sample regimes while
maintaining high TCBS. CROW also benefits from longer training, but its
budget-matched random target selection leads to less stable suppression,
especially when the selected modules do not coincide with the dominant
trigger-propagation pathway.

The \emph{middle}-trigger setting shows the clearest advantage for localization.
For $K=5$ and $K=7$, our 30-epoch detoxification reduces TMRR to nearly zero
with as few as $50$--$100$ training examples, while preserving strong TCBS. This
suggests that the proposed utility and redundancy criteria identify modules that
are not only individually influential but also complementary as a repair set.
By contrast, CROW-style consistency training reduces TMRR in some settings but
retains substantially higher malicious response rates in many matched-budget
comparisons, showing that a stronger repair objective cannot fully compensate
for untargeted module placement.

The \emph{end}-trigger setting is more competitive. CROW achieves strong
30-epoch performance in several settings, particularly when the sample budget is
moderate or large. This behavior is consistent with the intuition that a trigger
inserted close to the response boundary may rely on a shallower or more local
computation, making it easier for a consistency-based objective to disrupt.
Nevertheless, our method remains competitive and reaches zero TMRR in multiple
$K=5$ and $K=7$ settings, indicating that the proposed localization mechanism is
effective across all trigger positions rather than being specialized to a single
prompt template.

Across the three trigger placements, TCBS remains high in most settings where
TMRR is substantially reduced. This is important because it shows that the
detoxified models are not simply suppressing generation or drifting away from
the original benign behavior. Instead, the triggered outputs become closer to the
clean-response trajectory of the poisoned model, which is the intended behavior
of targeted post hoc detoxification.

\section{Discussion}

The results suggest that post hoc LLM backdoor removal is both an optimization
problem and a structural identification problem. Consistency-based objectives
such as CROW can improve internal stability under perturbation, but their
effectiveness under a limited adapter budget depends on whether the trainable
modules are placed on the relevant trigger-propagation pathway. Our method
addresses this placement problem directly by using activation patching and
curvature response to identify high-impact, nonredundant modules before applying
low-rank detoxification.

This distinction is important in practical deployment settings where broad
fine-tuning may be undesirable due to computational cost, limited clean data, or
risk of degrading benign capabilities. By freezing the poisoned model and
training only small additive adapters on selected modules, the proposed approach
offers a localized intervention that can suppress trigger-conditioned behavior
while preserving the model's clean-response trajectory.

The trigger-position results further indicate that backdoor pathways are not
uniform across prompt configurations. Beginning and middle triggers benefit most
from mechanistic localization, suggesting that their effects propagate through
more distributed internal computations. End triggers are more competitive for
CROW-style repair, likely because they occur closer to the response boundary and
may rely on shallower computations. This variation supports the need for defenses
that analyze internal propagation rather than assuming a fixed repair location.

Several limitations remain. The experiments use a controlled trigger, one base
model family, and aligned clean/triggered prompt pairs. Future work should
evaluate larger models, more diverse triggers and malicious objectives,
multi-trigger attacks, and settings where the trigger is only partially known or
must be inferred. It would also be valuable to study whether the same localization
signals can guide other interventions, such as pruning, model editing, or adapter
merging.

Overall, the findings support the view that LLM backdoor detoxification can be
treated as localized structural repair: identify where the trigger-conditioned
computation is mediated, then apply a compact correction only at those sites.
\bibliographystyle{IEEEtran}
\bibliography{references}

@inproceedings{xu2024instructions,
  title     = {Instructions as Backdoors: Backdoor Vulnerabilities of Instruction Tuning for Large Language Models},
  author    = {Xu, Jiashu and Ma, Mingyu and Wang, Fei and Xiao, Chaowei and Chen, Muhao},
  booktitle = {Proceedings of the 2024 Conference of the North American Chapter of the Association for Computational Linguistics: Human Language Technologies (Volume 1: Long Papers)},
  pages     = {3111--3126},
  year      = {2024}
}

@article{hubinger2024sleeper,
  title   = {Sleeper Agents: Training Deceptive LLMs that Persist Through Safety Training},
  author  = {Hubinger, Evan and Denison, Carson and Mu, Jesse and Lambert, Mike and Tong, Meg and MacDiarmid, Monte and Lanham, Tamera and Ziegler, Daniel M. and Maxwell, Tim and Cheng, Newton and Jermyn, Adam and Askell, Amanda and Radhakrishnan, Ansh and Anil, Cem and Duvenaud, David and Ganguli, Deep and Barez, Fazl and Clark, Jack and Ndousse, Kamal and Sachan, Kshitij and Sellitto, Michael and Sharma, Mrinank and DasSarma, Nova and Grosse, Roger and Kravec, Shauna and Bai, Yuntao and Witten, Zachary and Favaro, Marina and Brauner, Jan and Karnofsky, Holden and Christiano, Paul and Bowman, Samuel R. and Graham, Logan and Kaplan, Jared and Mindermann, S{"o}ren and Greenblatt, Ryan and Shlegeris, Buck and Schiefer, Nicholas and Perez, Ethan},
  journal = {arXiv preprint arXiv:2401.05566},
  year    = {2024}
}

@inproceedings{li2024backdoorllm,
  title     = {BackdoorLLM: A Comprehensive Benchmark for Backdoor Attacks and Defenses on Large Language Models},
  author    = {Li, Yige and Huang, Hanxun and Zhao, Yunhan and Ma, Xingjun and Sun, Jun},
  booktitle = {Advances in Neural Information Processing Systems Datasets and Benchmarks Track},
  year      = {2025}
}

@inproceedings{li2024simulate,
  title     = {Simulate and Eliminate: Revoke Backdoors for Generative Large Language Models},
  author    = {Li, Haoran and Chen, Yulin and Zheng, Zihao and Hu, Qi and Chan, Chunkit and Liu, Heshan and Song, Yangqiu},
  booktitle = {Proceedings of the AAAI Conference on Artificial Intelligence},
  year      = {2025}
}

@inproceedings{meng2022rome,
  title     = {Locating and Editing Factual Associations in {GPT}},
  author    = {Meng, Kevin and Bau, David and Andonian, Alex and Belinkov, Yonatan},
  booktitle = {Advances in Neural Information Processing Systems},
  volume    = {35},
  pages     = {17359--17372},
  year      = {2022}
}

@inproceedings{meng2023memit,
  title     = {Mass-Editing Memory in a Transformer},
  author    = {Meng, Kevin and Sharma, Arnab Sen and Andonian, Alex and Belinkov, Yonatan and Bau, David},
  booktitle = {The Eleventh International Conference on Learning Representations},
  year      = {2023}
}

@article{wang2024ke_survey,
  title   = {Knowledge Editing for Large Language Models: A Survey},
  author  = {Wang, Song and Zhu, Yaochen and Liu, Haochen and Zheng, Zaiyi and Chen, Chen and Li, Jundong},
  journal = {ACM Computing Surveys},
  year    = {2024}
}

@inproceedings{gu2024model,
  title     = {Model Editing Harms General Abilities of Large Language Models: Regularization to the Rescue},
  author    = {Gu, Jia-Chen and Xu, Hao-Xiang and Ma, Jun-Yu and Lu, Pan and Ling, Zhen-Hua and Chang, Kai-Wei and Peng, Nanyun},
  booktitle = {Proceedings of the 2024 Conference on Empirical Methods in Natural Language Processing},
  pages     = {16801--16819},
  year      = {2024}
}

@inproceedings{wang2024wise,
  title     = {WISE: Rethinking the Knowledge Memory for Lifelong Model Editing of Large Language Models},
  author    = {Wang, Peng and Li, Zexi and Zhang, Ningyu and Xu, Ziwen and Yao, Yunzhi and Jiang, Yong and Xie, Pengjun and Huang, Fei and Chen, Huajun},
  booktitle = {Advances in Neural Information Processing Systems},
  year      = {2024}
}

@article{heimersheim2024activation,
  title   = {How to Use and Interpret Activation Patching},
  author  = {Heimersheim, Stefan and Nanda, Neel},
  journal = {arXiv preprint arXiv:2404.15255},
  year    = {2024}
}

@inproceedings{lamparth2024analyzing,
  title     = {Analyzing and Editing Inner Mechanisms of Backdoored Language Models},
  author    = {Lamparth, Max and Reuel, Anka},
  booktitle = {Proceedings of the 2024 ACM Conference on Fairness, Accountability, and Transparency},
  pages     = {2362--2373},
  year      = {2024}
}

@inproceedings{hu2022lora,
  title     = {LoRA: Low-Rank Adaptation of Large Language Models},
  author    = {Hu, Edward J. and Shen, Yelong and Wallis, Phillip and Allen-Zhu, Zeyuan and Li, Yuanzhi and Wang, Shean and Wang, Lu and Chen, Weizhu},
  booktitle = {International Conference on Learning Representations},
  year      = {2022}
}

@inproceedings{zhang2020bertscore,
  title     = {BERTScore: Evaluating Text Generation with {BERT}},
  author    = {Zhang, Tianyi and Kishore, Varsha and Wu, Felix and Weinberger, Kilian Q. and Artzi, Yoav},
  booktitle = {International Conference on Learning Representations},
  year      = {2020}
}

@misc{mental_health,
  author       = {{Amod}},
  title        = {Mental Health Counseling Conversations},
  year         = {2025},
  howpublished = {Hugging Face dataset},
  note         = {\url{https://huggingface.co/datasets/Amod/mental_health_counseling_conversations}}
}

@inproceedings{min2025crow,
  title     = {CROW: Eliminating Backdoors from Large Language Models via Internal Consistency Regularization},
  author    = {Min, Nay Myat and Pham, Long H. and Li, Yige and Sun, Jun},
  booktitle = {Proceedings of the 42nd International Conference on Machine Learning},
  year      = {2025}
}

@article{yan2023cube,
  title   = {{CUBE}: A Black-Box Backdoor Defense via Clean Unlearning},
  author  = {Yan, Jun and Yadav, Vikas and Li, Shiyang and Chen, Lichang and Tang, Zheng and Wang, Hai and Srinivasan, Vijay and Ren, Xiang and Jin, Hongxia},
  journal = {arXiv preprint arXiv:2207.10348},
  year    = {2023}
}

@inproceedings{wu2025graceful,
  title     = {Gracefully Filtering Backdoor Samples for Generative Language Models},
  author    = {Wu, Zhaohan and others},
  booktitle = {Proceedings of the 31st International Conference on Computational Linguistics},
  year      = {2025}
}

@inproceedings{qi2021onion,
  title     = {{ONION}: A Simple and Effective Defense Against Textual Backdoor Attacks},
  author    = {Qi, Fanchao and Chen, Yangyi and Li, Mukai and Yao, Yuan and Liu, Zhiyuan and Sun, Maosong},
  booktitle = {Proceedings of the 2021 Conference on Empirical Methods in Natural Language Processing},
  pages     = {9558--9566},
  year      = {2021}
}

@article{mo2025testtime,
  title   = {Test-time Backdoor Mitigation for Black-Box Large Language Models with Defensive Demonstrations},
  author  = {Mo, Wenjie and others},
  journal = {arXiv preprint arXiv:2501.14725},
  year    = {2025}
}

@inproceedings{liu2018finepruning,
  title     = {Fine-Pruning: Defending Against Backdooring Attacks on Deep Neural Networks},
  author    = {Liu, Kang and Dolan-Gavitt, Brendan and Garg, Siddharth},
  booktitle = {International Symposium on Research in Attacks, Intrusions, and Defenses},
  pages     = {273--294},
  year      = {2018},
  publisher = {Springer}
}

\end{document}